\documentclass[aps,prl,twocolumn,superscriptaddress]{revtex4-1} 
\usepackage{graphicx}

\usepackage{amsmath,bm}

\begin{document}
\title{A Near-Ideal Molecule-Based Haldane Spin-Chain}

\author{Robert C. Williams}
\author{William J. A. Blackmore}
\author{Samuel P. M. Curley}
\author{Martin R. Lees}
\affiliation{Department of Physics, University of Warwick, Gibbet Hill Road, Coventry CV4 7AL, UK}
\author{Serena M. Birnbaum}
\author{John Singleton}
\affiliation{National High Magnetic Field Laboratory, Los Alamos National Laboratory, Los Alamos, New Mexico 87545, USA}
\author{Benjamin M. Huddart}
\author{Thomas J. Hicken}
\author{Tom Lancaster}
\affiliation{Department of Physics, Durham University, South Road, Durham DH1 3LE, UK}
\author{Stephen J. Blundell}
\affiliation{Department of Physics, Clarendon Laboratory, Oxford University, Parks Road, Oxford OX1 3PU, UK}
\author{Fan Xiao}
\affiliation{Department of Chemistry and Biochemistry, University of Bern, Freiestrasse 3, CH-3012 Bern, Switzerland}
\affiliation{Laboratory for Neutron Scattering and Imaging, Paul Scherrer Institut, CH-5232 Villigen PSI, Switzerland}
\author{Andrew Ozarowski}
\affiliation{National High Magnetic Field Laboratory, Florida State University, Tallahassee, Florida 32310, USA}
\author{Francis L. Pratt}
\author{David J. Voneshen}
\affiliation{ISIS Facility, STFC Rutherford Appleton Laboratory, Chilton,  Oxon OX11 0QX, UK}
\author{Zurab Guguchia}
\author{Christopher Baines}
\affiliation{Laboratory for Muon Spin Spectroscopy, Paul Scherrer Institut, CH-5232 Villigen PSI, Switzerland}
\author{John A. Schlueter}
\affiliation{Materials Science Division, Argonne National Laboratory, Argonne, Illinois 60439, USA}
\affiliation{Division of Materials Research, National Science Foundation, 2415 Eisenhower Ave, Alexandria, Virginia 22314, USA}
\author{Danielle Y. Villa}
\author{Jamie L. Manson}
\affiliation{Department of Chemistry and Biochemistry, Eastern Washington University, 226 Science, Cheney, Washington 99004, USA}
\author{Paul A. Goddard}
\email{p.goddard@warwick.ac.uk}
\affiliation{Department of Physics, University of Warwick, Gibbet Hill Road, Coventry CV4 7AL, UK}
\begin{abstract}

The molecular coordination complex NiI$_2$(3,5-lut)$_4$ [where (3,5-lut) $=$ (3,5-lutidine)  $=$ (C$_7$H$_9$N)] has been synthesized and characterized by several techniques including synchrotron X-ray diffraction, ESR, SQUID magnetometry, pulsed-field magnetization, inelastic neutron scattering and muon spin relaxation. Templated by the configuration of 3,5-lut ligands the molecules pack in-registry with the Ni--I$\cdots$I--Ni chains aligned along the $c$--axis. This arrangement leads to through-space I$\cdots$I magnetic coupling which is directly measured for the first time in this work. The net result is a near-ideal realization of the $S = 1$ Haldane chain with $J = 17.5~\rm{K}$ and energy gaps of $\Delta^{\parallel} = 5.3~{\rm K}$ $\Delta^{\perp} =7.7~{\rm K}$, split by the easy-axis single-ion anisotropy  $D=-1.2~{\rm K}$.
The ratio $D/J = -0.07$ affords one of the most isotropic Haldane systems yet discovered, while the ratio $\Delta_0/J = 0.40(1)$ (where $\Delta_0$ is the average gap size) is close to its ideal theoretical value, suggesting a very high degree of magnetic isolation of the spin chains in this material. The Haldane gap is closed by orientation-dependent critical fields $\mu_0H_{\rm c}^{\parallel} = 5.3~\rm{T}$ and $\mu_0H_{\rm c}^{\perp} =  4.3~\rm{T}$, which are readily accessible experimentally and permit investigations across the entirety of the Haldane phase, with the fully polarized state occurring at $\mu_0 H_{\rm s}^{\parallel}=46.0~\rm{T}$ and $\mu_0 H_{\rm s}^{\perp}=50.7~\rm{T}$. The results are explicable within the so-called fermion model, in contrast to other reported easy-axis Haldane systems. Zero-field magnetic order is absent down to $20~{\rm mK}$ and emergent end-chain effects are observed in the gapped state, as evidenced by detailed low-temperature measurements.

\end{abstract}

\pacs{}
\maketitle

\section{Introduction}

{\textbf{Background.} Molecule-based magnets have proved a highly successful avenue in achieving desired low-dimensional geometries \cite{batten,blundell2007,goddard2012}. Several classes of materials which host quantum-disordered non-magnetic ground states can undergo magnetic field and pressure driven quantum phase transitions (QPTs), where the magnetic ground states undergo a dramatic reorganisation and physical properties exhibit characteristic behavior \cite{sachdev2011,zapf2014}.

An important asset of molecule-based systems is the modest energy scale of their exchange interaction strengths, which makes their QPTs experimentally accessible, thus providing a vital opportunity to test the predictions of theory and simulations. Invaluable insight has been afforded by the intensive experimental study of prototypical realizations of models including: the ideal $S=1/2$ chain material Cu(C$_4$H$_4$N$_2$)(NO$_3$)$_2$ [\citenum{hammar1999}]; the strong-leg and strong-rung two-leg spin-ladder systems (C$_7$H$_{10}$N)$_2$CuBr$_4$ (DIMPY) [\citenum{hong2010}] and (C$_5$H$_{12}$N)$_2$CuBr$_4$ (BPCB) [\citenum{watson2001}], respectively; the $S=1/2$ dimer compound [Cu(pyz)(gly)](ClO$_4$) [\citenum{lancaster2014}]; the large-$D$ system NiCl$_2$−4SC(NH$_2$)$_2$ (DTN) [\citenum{zapf2006}]; and the spin-liquid compound (C$_4$H$_{12}$N$_2$)Cu$_2$Cl$_6$ (PHCC) [\citenum{stone2006}]. Of note is that several of these halide containing materials rely on through-space magnetic couplings; i.e.\ two-halide exchange, rather than that offered by bridging ligands. For the first time, we present a novel material of this kind supported by non-covalent I$\cdots$I interactions. 

Pioneering work by Haldane demonstrated that Heisenberg antiferromagnetic (AFM) chains of integer spins host a disordered ground state which is topologically distinct from the well-known, spin-half analogue (the Tomonaga-Luttinger liquid) \cite{haldane1983,haldane1983a}. The significance of this was recognized via the award of the 2016 Nobel Prize in Physics. The Haldane phase is protected by its eponymous energy gap, above which excitations propagate. Fractional excitations manifesting as $S=1/2$ degrees of freedom exist at the end-chains,  due to the symmetry-protected bulk topological phase. This phenomenon arises naturally out of the valence bond solid model \cite{affleck1987}, and has been experimentally observed in several Haldane compounds \cite{hagiwara1990,glarum1991,avenel1992}. Higher concentrations of $S=1/2$ edge-state excitations may be induced by breaking the chains via the introduction of non-magnetic sites, which has been shown to result in magnetic ordering in PbNi$_2$V$_2$O$_8$ due to the almost critical inter-chain exchange coupling \cite{uchiyama1999,zheludev2000}.

It has been demonstrated that extending the Hamiltonian of the AFM $S=1$ chain to include single-ion anisotropy (SIA) and inter-chain interactions gives rise to a rich phase diagram \cite{sakai1990,wierschem2014,wierschem2014a}. The resultant Hamiltonian is given by
\begin{multline}
\hat{\mathcal{H}} = D\sum_{i}(\hat{\bm{S}}^{z}_{i})^{2} 
+ J \sum_{\langle i,j\rangle}\hat{\bm{S}}_{i}\cdot\hat{\bm{S}}_{j} 
+ J_{\perp}\sum_{\langle i,j'\rangle}\hat{\bm{S}}_{i}\cdot\hat{\bm{S}}_{j'}
\\+ \mu _{\rm{B}}\mu _{0}\sum_{i}\bm{H}\cdot \bm{g}\cdot \hat{\bm{S}}_{i},
\label{eq:ham}
\end{multline}
\noindent
where $J$ and $J_{\perp}$ are the dominant intra-chain and weaker inter-chain exchange interactions between $S=1$ moments, respectively. Here, angular brackets denote sums taken over unique pairs of nearest-neighbor spins, primed indices refer to spins on adjacent chains, $D$ is the axial SIA parameter and the final term is the usual Zeeman interaction for an external magnetic field, $\bm{H}$. The ensuing phase diagram has recently been extended to consider the rhombic SIA parameter, $E$ [\citenum{tzeng2017}].

In order to exploit the interplay between crystal structure and magnetic behavior in  $S=1$ systems, and ultimately tune and control the parameters $J$ and $D$, we have synthesized and characterized a series of Ni(II)-containing materials. Different bridging and non-bridging ligands enable the exploration of low-dimensional phase diagrams, resulting in dramatically different magnetic behavior \cite{blackmore,liu2016,blackmore2019}. In this work we report the discovery and characterization of a $S=1$ molecule-based AFM chain NiI$_2$(3,5-lut)$_4$ where (3,5-lut) $=$ (3,5-lutidine)  $=$ (C$_7$H$_9$N). Magnetometry measurements reveal the AFM intra-chain coupling $J=17.5(2)~{\rm K}$, which together with the axial SIA parameter $D=-1.2(1)~{\rm K}$ results in two triplet energy gaps $\Delta^{\parallel} = 0.46(1)~{\rm meV} = 5.3(1)~{\rm K}$ and $\Delta^{\perp} = 0.66(1)~{\rm meV}=7.7(1)~{\rm K}$, as observed via inelastic neutron scattering (INS). The Haldane gap may be closed by the anisotropic critical magnetic fields $\mu_0 H_{\rm c}^{\parallel}=5.3(1)~{\rm T}$ and  $\mu_0 H_{\rm c}^{\perp}=4.3(1)~{\rm T}$, and the system is saturated at $\mu_0 H_{\rm s}^{\parallel}=46.0(4)~{\rm T}$ and  $\mu_0 H_{\rm s}^{\perp}=50.7(8)~{\rm T}$. The ratio $D/J = -0.07(1)$ makes this the among the most isotropic Haldane chains reported to date, and the modest energy scales lead to experimentally accessible critical fields, enabling the use of superconducting magnets found at beamline facilities to explore the entire Haldane phase. To this end, we performed the first transverse-field muon spin relaxation ($\mu^{+}$SR) study of the field-driven closure of the Haldane gap, while complementary zero-field $\mu^{+}$SR measurements confirm the absence of zero-field magnetic ordering down to $T = 20~{\rm mK}$.

For the isotropic Haldane phase with $D=0$, analytical and numerical approaches  postulate two- and three-particle continua in the zero-field magnetic excitation spectra at the Brillouin zone boundary ($k=\pi$) and center ($k=0$), respectively (see Ref.~\cite{rahnavard2015} and references therein). These continua augment the well-established lowest-lying ``single-magnon'' mode defining the Haldane gap $\Delta$. However, their existence has not yet been demonstrated experimentally \cite{ma1992,zaliznyak2001,kenzelmann2001}. A nearly isotropic Haldane chain material such as NiI$_2$(3,5-lut)$_4$ is therefore highly desirable to help verify this prediction. NiI$_2$(3,5-lut)$_4$ also affords the possibility for controlled introduction of both bond and site disorder, which makes this compound a model system for investigations into the physics of the Haldane phase.

\textbf{Single Ion Anistropy in the Haldane Phase.} In zero-field, the presence of axial SIA serves to lift the degeneracy of the excited triplet states above the Haldane ground state, resulting in energy levels given by \cite{golinelli1992}
\begin{align}\label{eq:ZF}
\Delta^{\parallel}  = \Delta_0 + 1.41D, && \Delta^{\perp}  = \Delta_0 - 0.57D,
\end{align}
for the singlet longitudinally polarized excitation mode  (denoted $\parallel$, corresponding to the $S_z = 0$ triplet state) and the transverse-polarized doublet  ($\perp$, corresponding to the $S_z = \pm 1$ triplet states). Longitudinal and transverse is defined with respect to the unique $z$ direction, which here coincides with the chain axis $c$. The intrinsic Haldane gap $\Delta_0 = 0.41 J$, depends solely on the intra-chain exchange coupling \cite{nightingale1986,white1993,golinelli1994,todo2001,white2008}. This energy-level arrangement may be compared to the single-ion equivalent for isolated $S=1$ moments subject to axial SIA, where the effective SIA parameter $D_{\rm eff}$ is given by
\begin{equation}\label{eq:Deff}
D_{\rm eff} = \Delta^{\perp}-\Delta^{\parallel} = -1.98D.
\end{equation}
Note $D_{\rm eff}$ and $D$ have opposite signs.

There are several different quantum-field-theoretical models which describe the evolution of these energy levels under the application of magnetic fields. These models yield anisotropic values of the critical fields at which the Zeeman interaction drives a triplet state lower in energy than the Haldane ground state.

The field-dependent energy levels of the `fermion' (and equivalent `perturbative') model \cite{affleck1992,golinelli1993,regnault1993,tsvelik1990} map directly onto the isolated $S=1$ single-ion description \cite{abragam}, with effective axial SIA term $D_{\rm eff}$ given by Equation~\ref{eq:Deff}. The fermion model predicts the following critical fields
\begin{align}\label{eq:ferm}
g_{\parallel} \mu_{\rm B} \mu_0 H_{\rm c}^{\parallel} = \Delta^{\perp}, && g_{\perp} \mu_{\rm B} \mu_0 H_{\rm c}^{\perp} = \sqrt{\Delta^{\parallel} \Delta^{\perp}},
\end{align}
for fields parallel and perpendicular to the unique axis.

The `boson' (and equivalent `macroscopic') model \cite{affleck1992,farutin2007} differs from the fermion model for fields perpendicular to the unique axis, particularly in the region close to the field-driven closure of the Haldane gap. The boson model predicts the following anisotropic critical fields
\begin{align}\label{eq:bos}
g_{\parallel} \mu_{\rm B} \mu_0 H_{\rm c}^{\parallel} = \Delta^{\perp}, && g_{\perp} \mu_{\rm B} \mu_0 H_{\rm c}^{\perp} = \Delta^{\parallel}.
\end{align}
Note that for the case of a negative SIA parameter $D<0$ (easy-axis), then $\Delta^{\parallel} < \Delta^{\perp} $ and therefore both the fermion and boson models predict $H_{\rm c}^{\parallel} > H_{\rm c}^{\perp}$ (and vice-versa for the easy-plane case).

The fermion model has been shown to consistently describe the behavior of the easy-plane compounds NENP [\citenum{regnault1994}] and NDMAP [\citenum{zheludev2003}], as it is able to simultaneously account for the anisotropic critical fields and the triplet energy levels. In contrast, the boson model has been shown to successfully describe the behavior of the two easy-axis Haldane compounds PbNi$_2$V$_2$O$_8$ and SrNi$_2$V$_2$O$_8$ [\citenum{smirnov2008,bera2015,bera2015a}]. It is believed that the success of the boson model for these compounds is due to their easy-axis anisotropy and critical inter-chain coupling interactions ($J_{\perp}$), which place them close to the phase boundary between the Haldane and ordered Ising AFM phases. 

\section{Results}

\begin{figure}[b]
  \includegraphics[width=\columnwidth]{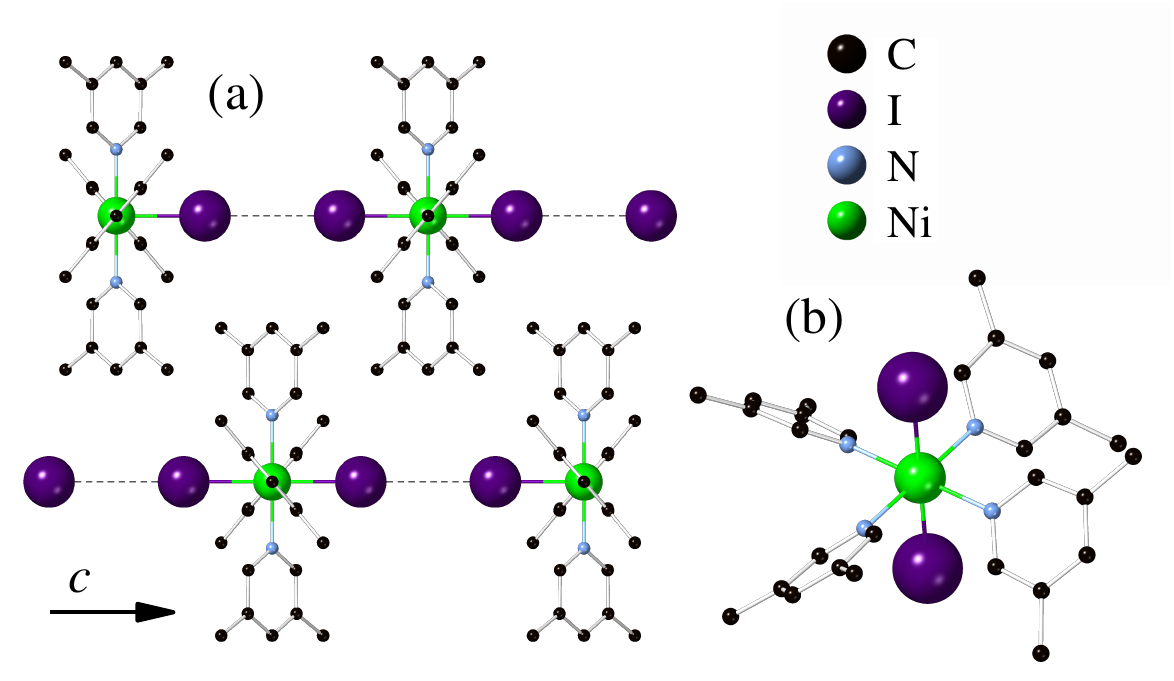}
  \caption{(a) The chain structure of NiI$_2$(3,5-lut)$_4$, viewed along the $[110]$ direction, normal to the chain axis $\bm{c}$ (H atoms are omitted for clarity). (b) The local crystal environment of each Ni(II) ion.}
  \label{fig:structure}
\end{figure}

\textbf{Structural Determination.} 
Synchrotron single-crystal X-ray diffraction measurements were performed to solve the crystal structure of NiI$_2$(3,5-lut)$_4$ at $T = 100(2)~{\rm K}$. NiI$_2$(3,5-lut)$_4$ crystallizes in the space group $P4/nnc$, and comprises linear Ni--I$\cdots$I--Ni chains running parallel to the $c$-axis, as shown in Figure~\ref{fig:structure}(a). The intra-chain Ni--Ni separation is $c=9.9783(2)$~\AA{}, where the magnetic superexchange interaction $J$ is mediated via the two iodine ions, and the linear configuration suggests AFM interactions according to the Goodenough-Kanamori rules \cite{Goodenough1963,Kanamori1959,Anderson1963}.

The local crystal environment of each Ni ion is a tetragonally elongated NiI$_2$N$_4$ octahedron, shown in Figure~\ref{fig:structure}(b), where the unique axial direction coincides with the chain axis $\bm{c}$. The Ni--N and Ni--I bond lengths are 2.123 and 2.833~\AA{}, respectively. The chains running parallel to $\bm{c}$ lie on the axes of four-fold rotational symmetry operators, which preclude the existence of rhombic SIA, and we therefore expect the magnetic behavior of NiI$_2$(3,5-lut)$_4$ to be described by the Hamiltonian in Equation~\ref{eq:ham}.

The non-bridging lutidine ligands serve to separate adjacent chains within the $ab$-plane, which are offset by $\bm{c}/2$. The shortest Ni--Ni distance is actually between ions on neighboring chains, however, there is no clear exchange pathway in this direction. The rings of the lutidine ligands are canted by $41.57^{\circ}$ with respect to the $ab$-plane, forming a propeller-like arrangement about the chain axes. 

Not only do the methyl-substituents of the lutidine ligands in NiI$_2$(3,5-lut)$_4$ create large inter-chain separations but they influence the crystal packing and prompt proper alignment of Ni--I$\cdots$I--Ni chains. While NiI$_2$(pyridine)$_4$ is known \cite{Hamm1973}, the molecules pack in such a way that potential I$\cdots$I interactions are negated, making NiI$_2$(3,5-lut)$_4$ particularly novel. It is now possible to directly measure the magnetic exchange strength propagated by I$\cdots$I couplings.

\textbf{Magnetometry.} 
Magnetic susceptibility measurements were performed on a powder sample of NiI$_2$(3,5-lut)$_4$ using a SQUID magnetometer, with results shown in Figure~\ref{fig:squid}(a). Upon cooling, a broad peak in the molar susceptibility $\chi_{\rm m} (T)$ is visible at around $20~{\rm K}$ which indicates the build-up of short-range correlations and is typical for an AFM chain. As temperature is decreased further, a local minimum is reached at around $2~{\rm K}$, due to a paramagnetic tail which dominates the magnetic response as the sample is cooled toward zero temperature. Importantly, there is no evidence in $\chi_{\rm m}(T)$ for a transition to magnetic long range order (LRO) down to the lowest measured temperature of $0.47~{\rm K}$. In fact, as will be discussed later, our muon spin relaxation study demonstrates that in zero-field NiI$_2$(3,5-lut)$_4$ does not undergo magnetic ordering for temperatures down to $T = 20~{\rm mK}$, indicating a small ratio $|J_{\perp}|/J$. Taken together with the small ratio $|D|/J$ revealed by the INS results, discussed below, this places NiI$_2$(3,5-lut)$_4$ in the gapped Haldane region of the theoretically derived phase diagram \cite{sakai1990,wierschem2014}. The paramagnetic upturn in $\chi_{\rm m} (T)$ is therefore ascribed to the emergence of $S=1/2$ degrees of freedom at the chain-ends within the Haldane phase, as has been observed in susceptibility studies of related Haldane systems, for example 
NENP [\citenum{avenel1992}].

\begin{figure}[t]
  \includegraphics[width=\columnwidth]{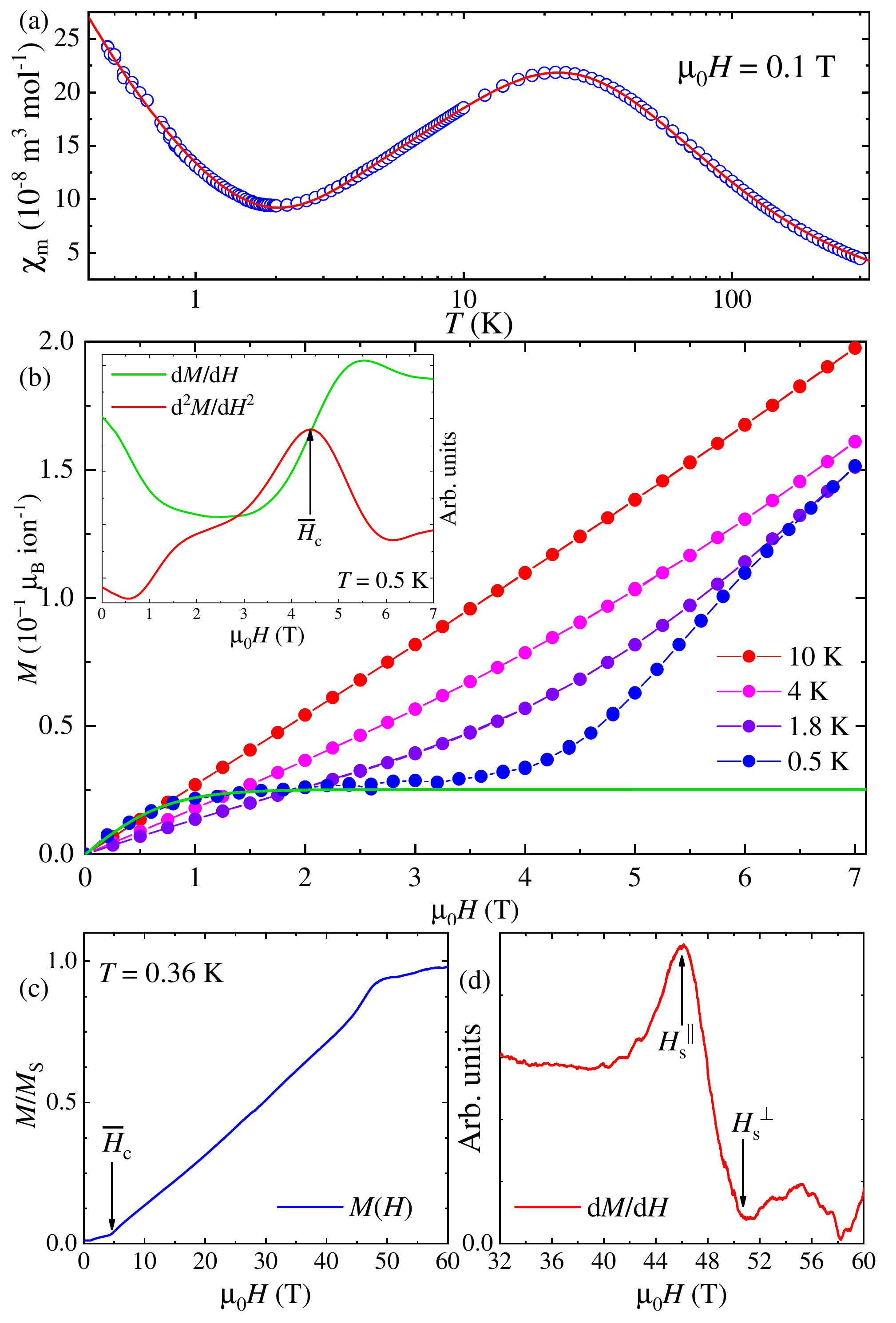}
  \caption{Powder magnetometry results for NiI$_2$(3,5-lut)$_4$. (a) Magnetic susceptibility data (circles), where the solid line represents a fit to the data over the entire measured temperature range using Equation~\ref{eq:law}. (b) Quasi-static magnetization data, where the lowest temperature ($T = 0.5~{\rm K}$) data have been fit to Equation~\ref{eq:mag} for the field range $\mu_0 H \leq 2 ~{\rm T}$ (solid green line). (Inset) The first and second derivatives of the low-temperature $M(H)$ data, with respect to magnetic field. (c) Pulsed-field magnetization data and (d) their derivative, measured at $T = 0.36~{\rm K}$.
  }
  \label{fig:squid}
\end{figure}

Defect sites (with fraction $\rho$) each induce two $S=1/2$ end-chain moments, and are expected to arise due to lattice defects and crystallite boundaries within the powder sample. The temperature-dependent molar susceptibility data may therefore be modeled using the sum of two components
\begin{equation}
\chi_{\rm m} = (1- \rho ) \chi_{\rm chain} + 2\rho \chi_{1/2} ,
\label{eq:law}
\end{equation}
where $\chi_{\rm chain}$ is the susceptibility of an ideal isotropic Haldane chain \cite{law2013,bera2013} and the end-chain contribution is captured by $\chi_{1/2}$, which is the usual Curie-Weiss expression for paramagnetic $S=1/2$ moments. 

Despite the isotropic approximation, this model is seen to account for the $\chi_{\rm m}(T)$ data extremely well across the entire measured temperature range ($0.47 \leq T \leq 300 ~{\rm K}$), as shown in Figure~\ref{fig:squid}(a). The extracted fit parameters yield the fraction of defect sites $\rho = 1.7(1)\%$, plus estimates for the intra-chain exchange interaction strength $J = 18.33(7)~{\rm K}$ and the average Haldane gap $\Delta_0 = 7.44(1)~{\rm K}$. The resultant magnitude of the Haldane gap with respect to the intra-chain exchange strength $\Delta_0/J = 0.406(2)$ is found to be in excellent agreement with the theoretically expected value of $0.41$ for the ideal Haldane chain \cite{nightingale1986,white1993,golinelli1994,todo2001,white2008}. The fit value of the powder-average $g$-factor is $g=2.184(2)$.

Magnetization measurements were performed on a powder sample of NiI$_2$(3,5-lut)$_4$ in quasi-static magnetic fields $\mu_0 H \leq 7~{\rm T}$, with results shown in Figure~\ref{fig:squid}(b). A clear kink is visible in the data close to $4~{\rm T}$, which becomes more pronounced in lower temperature datasets. We ascribe this kink to the field-driven closure of the Haldane gap, as the Zeeman interaction drives the energy of a triplet excited state below that of the Haldane ground state \cite{renard2003}.  
The powder-average value of the critical field $\mu_0 \bar{H}_{\rm c}=4.4(2)~{\rm T}$ is obtained by locating the peak in ${\rm d}^2M/{\rm d}H^2$ at $T = 0.5~{\rm K}$, shown in the inset to Figure~\ref{fig:squid}(b), corresponding to the kink in the $M(H)$ magnetization data. 

Further evidence for the presence of the magnetically disordered Haldane phase for fields below $H_{\rm c}$, comes from the suppression of magnetization in this region as temperature is decreased from $T = 10~{\rm K}$ [Figure~\ref{fig:squid}(b)]. A magnetic response is still visible for low magnetic fields $\mu_0 H < 2~{\rm T}$ for $M(H)$ data measured at $T=0.5~{\rm K}$, which is due to the  paramagnetism of the end-chain $S=1/2$ moments already identified in the susceptibility data. These low-field magnetization data can be modeled using
\begin{equation}
M(H)=N_{1/2}M_{1/2}(H)+N_{1}M_{1}(H),
\label{eq:mag}
\end{equation}
where $N_S$ and $M_{S}(H)$ are the fractions and Brillouin responses, respectively, of paramagnetic moments with spin $S$. The fit yields $N_{1/2}=2.5(3)\%$, comparable to the fraction of end-chain moments $2\rho=3.4(2)\%$ obtained using the susceptibility data, while $N_{1}=0.0(2)\%$, further confirming the presence of fractional end-chain states arising out of the Haldane phase.

\begin{figure}[t]
\centering
 \includegraphics[width=0.9\columnwidth]{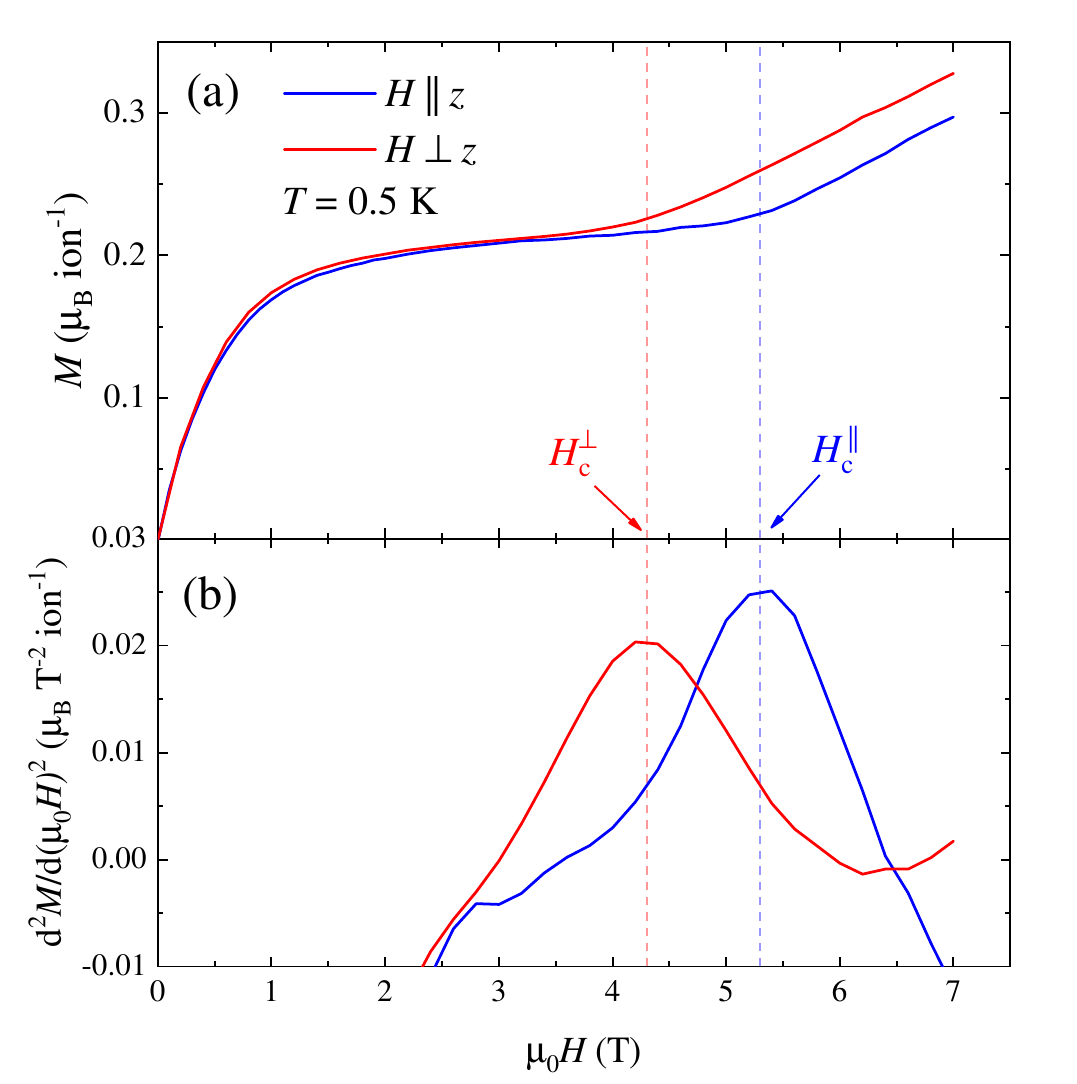}
 \caption{(a) Single-crystal magnetization measurements performed with magnetic fields parallel ($\parallel$) and perpendicular ($\perp$) to the unique local axis in NiI$_2$(3,5-lut)$_4$, which coincides with the chain direction. (b) The second derivative of magnetization with respect to field yields the anisotropic critical fields.}  \label{fig:SQUID2}
\end{figure}

In order to investigate the SIA-induced orientation-dependence of the critical magnetic fields, low-temperature ($T=0.5~{\rm K}$) magnetization measurements were made on a small single crystal of NiI$_2$(3,5-lut)$_4$ with magnetic fields oriented parallel ($\parallel$) and perpendicular ($\perp$) to the unique $c$-axis, with results shown in Figure~\ref{fig:SQUID2}. An isotropic  paramagnetic response is visible at low fields, consistent with the behaviour of paramagnetic $S=1/2$ moments; however, the subsequent upturn in $M(H)$ is clearly seen to differ for the two magnetic field orientations. By examining the second derivative ${\rm d}^2M/{\rm d}H^2$ [Figure~\ref{fig:SQUID2}(b)], the critical field values $\mu_0 H_{\rm c}^{\parallel}=5.3(1)~{\rm T}$ and  $\mu_0 H_{\rm c}^{\perp}=4.3(1)~{\rm T}$ are determined. The observation that $H_{\rm c}^{\parallel} > H_{\rm c}^{\perp}$ constitutes unambiguous evidence for non-zero easy-axis SIA  in this compound. 

Pulsed-field magnetization measurements were performed on a powder sample of NiI$_2$(3,5-lut)$_4$, with representative results shown in Figure~\ref{fig:squid}(c). As for the measurements under quasi-static fields, a kink is clearly visible where the Haldane gap is closed at $\bar{H}_{\rm c}$. As magnetic field is increased further, the magnetization response increases approximately linearly until close to the saturation field, where the  curvature is typical of a one-dimensional system \cite{bonner1964}. The peak in ${\rm d}^2M/{\rm d}H^2$ is used to determine the  value of the critical field $\mu_0 \bar{H}_{\rm c}=4.3(3)~{\rm T}$, which agrees closely with the value deduced from the SQUID magnetometry measurements. Using the result that $D<0$ in this system then, following the mean-field calculations in Ref.~\cite{brambleby2017}, the anisotropic saturation fields are found to be
\begin{align}\label{eq:sat}
g_{\parallel} \mu_{\rm B} \mu_0 H_{\rm s}^{\parallel} = 2(2J-|D|), && g_{\perp} \mu_{\rm B} \mu_0 H_{\rm s}^{\perp} = 2(2J+|D|),
\end{align}
where it has been assumed that $|J_{\perp}| \ll J$. As shown in Figure~\ref{fig:squid}(d), $\mu_0 H_{\rm s}^{\parallel}=46.0(4)~{\rm T}$ and $\mu_0 H_{\rm s}^{\perp}=50.7(8)~{\rm T}$ are readily identifiable as the beginning and end, respectively, of the decrease in ${\rm d}M/{\rm d}H$ accompanying the plateau in $M(H)$ at saturation. Taking the anisotropic $g$-factors $g_{\parallel,\perp}=2.13(1),2.19(1)$ from the ESR results discussed below, Equations~\ref{eq:sat} yield values for the SIA parameter $D=-1.2(3)~{\rm K}$ and intra-chain exchange coupling $J = 17.5(2)~{\rm K}$, which is in reasonable agreement with the estimate obtained from the fit to the susceptibility data.

\begin{figure}[b]
  \includegraphics[width=\columnwidth]{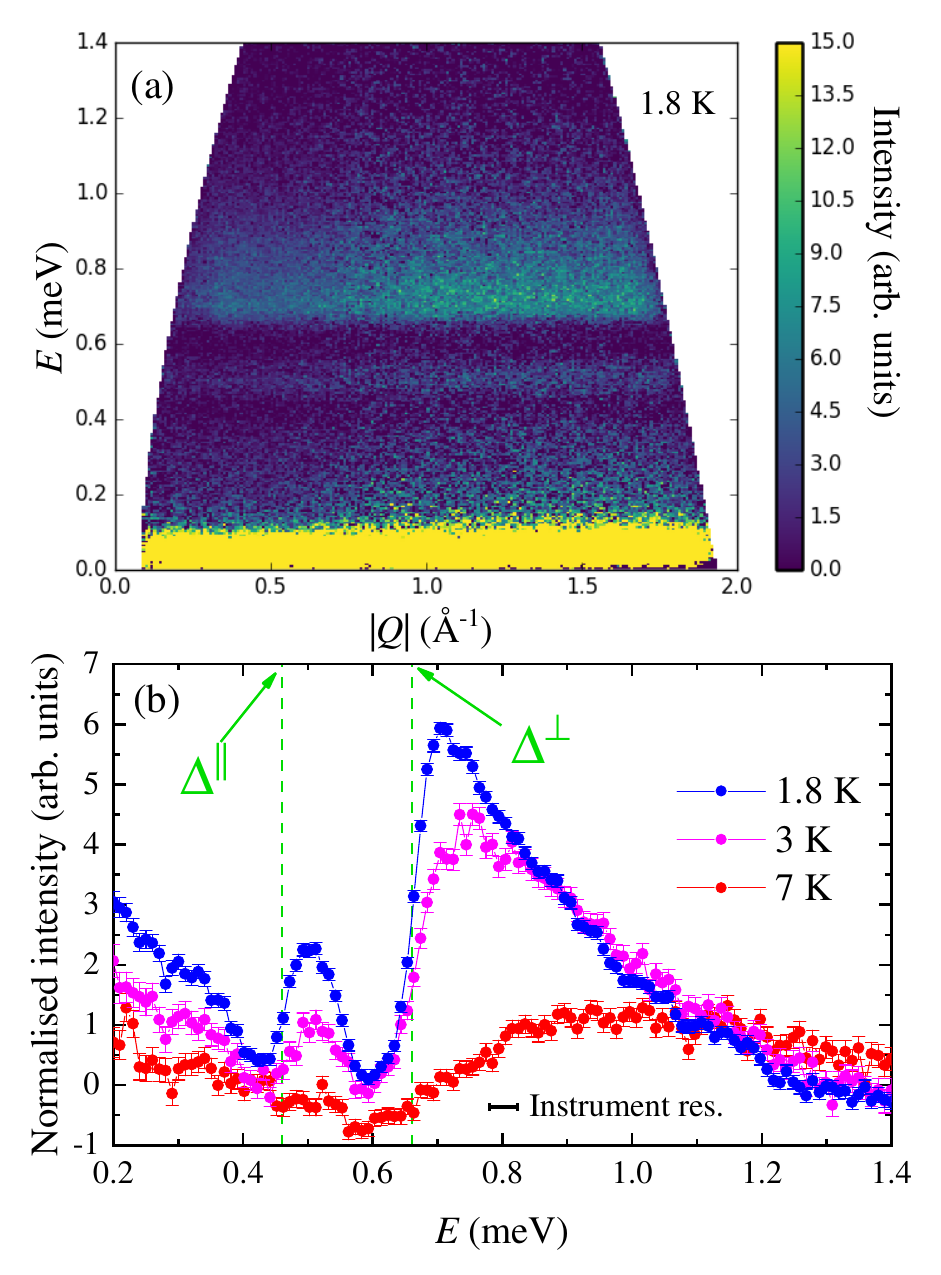}
  \caption{(a)  Time-of-flight INS data, presented as a ($|Q|,E$) heat-map for $T=1.8~{\rm K}$ and incident neutron energy $E_{\rm i}=2.2~{\rm meV}$, after background subtraction. (b) Representative energy cuts, obtained by integrating over the full measured range of $|Q|$, for the background-subtracted data at $T = 1.8$, $3$ and $7~{\rm K}$.}
  \label{fig:ins}
\end{figure}

\textbf{Inelastic Neutron Scattering.}
In order to examine the magnetic excitations of NiI$_2$(3,5-lut)$_4$ and resolve the SIA-split Haldane energy gaps in zero-field, INS measurements were performed on a powder sample using the LET instrument at ISIS, UK. Representative data collected at $T = 1.8~{\rm K}$ for incident neutron energy $E_{\rm i}=2.2~{\rm meV}$ are shown as a ($|Q|,E$) spectrum in Figure~\ref{fig:ins}(a), where $|Q|$ and $E$ denote momentum and energy transfers, respectively.
Corresponding $T=12~{\rm K}$ data are treated as a non-magnetic background, and have been subtracted from the data. The two components of the triplet excitations are visible as bands of excitations for energies above $E \approx 0.4$ and $0.6~{\rm meV}$. Intensity for energy transfers below $E \approx 0.2~{\rm meV}$ is attributed to the elastic and quasi-elastic contribution from nuclear scattering.

The Haldane gaps may be clearly quantified by inspecting an energy cut through the data: Figure~\ref{fig:ins}(b) shows cuts obtained by integrating over the full measured range of $|Q|$ for the background-subtracted data for $T=1.8$, $3$ and $7~{\rm K}$.  The SIA-split Haldane gaps are visible as the low-energy onsets of the bands of spin excitations,  where the large density of states at each excitation band minimum leads to a rapid increase in scattering intensity at energies centered around each gap energy \cite{mutka1991,zheludev2001a}. The excitation bands become more pronounced as temperature is lowered, confirming that these features originate in the magnetic contribution to the scattering.

Taking the midpoint of the low-energy onset of each peak yields values for the SIA-split Haldane gaps of $\Delta^{\parallel} = 0.46(1)~{\rm meV} = 5.3(1)~{\rm K}$ and $\Delta^{\perp} = 0.66(1)~{\rm meV}=7.7(1)~{\rm K}$, where labels $\parallel$ and $\perp$ are assigned using the conclusion that SIA in this system is easy-axis in nature, based on the  magnetization results discussed above. Using Equation~\ref{eq:ZF}, these gap values correspond to an easy-axis value of the SIA parameter $D = -1.2(1)~{\rm K}$, in excellent agreement with the estimate based on the anisotropic saturation fields $H_{\rm s}^{\parallel,\perp}$. The intrinsic Haldane gap of $\Delta_0=0.60(1)~{\rm meV} = 7.0(1)~{\rm K}$ predicts an intra-chain AFM exchange interaction $J = 17.0(2)$, where both of these values are in reasonable agreement with the result of the powder magnetometry measurements described above. The ratio of SIA to intra-chain exchange is therefore determined to be $D/J = -0.07(1)$ which, together with the absence of LRO for temperatures down to $T=20~{\rm mK}$ revealed using muon spin relaxation, places NiI$_2$(3,5-lut)$_4$ firmly in the Haldane phase at low temperatures, as per the phase diagram in Ref.~\cite{wierschem2014}.

\textbf{Muon Spin Relaxation.}
Zero-field muon spin relaxation ($\mu^{+}$SR) measurements were made on a powder sample of NiI$_2$(3,5-lut)$_4$  to search for LRO in the absence of any applied magnetic field. No oscillations were resolved in the muon spin polarization for temperatures down to $T = 20~{\rm mK}$, consistent with the absence of magnetic order. Instead, we observe exponential relaxation with very little temperature-dependence, implying dynamics in a disordered magnetic field distribution \cite{yaouanc} and no phase transition to an ordered state. This observation is consistent with the realization of a Haldane state.

\begin{figure}[t]
	\includegraphics[width=0.75\columnwidth]{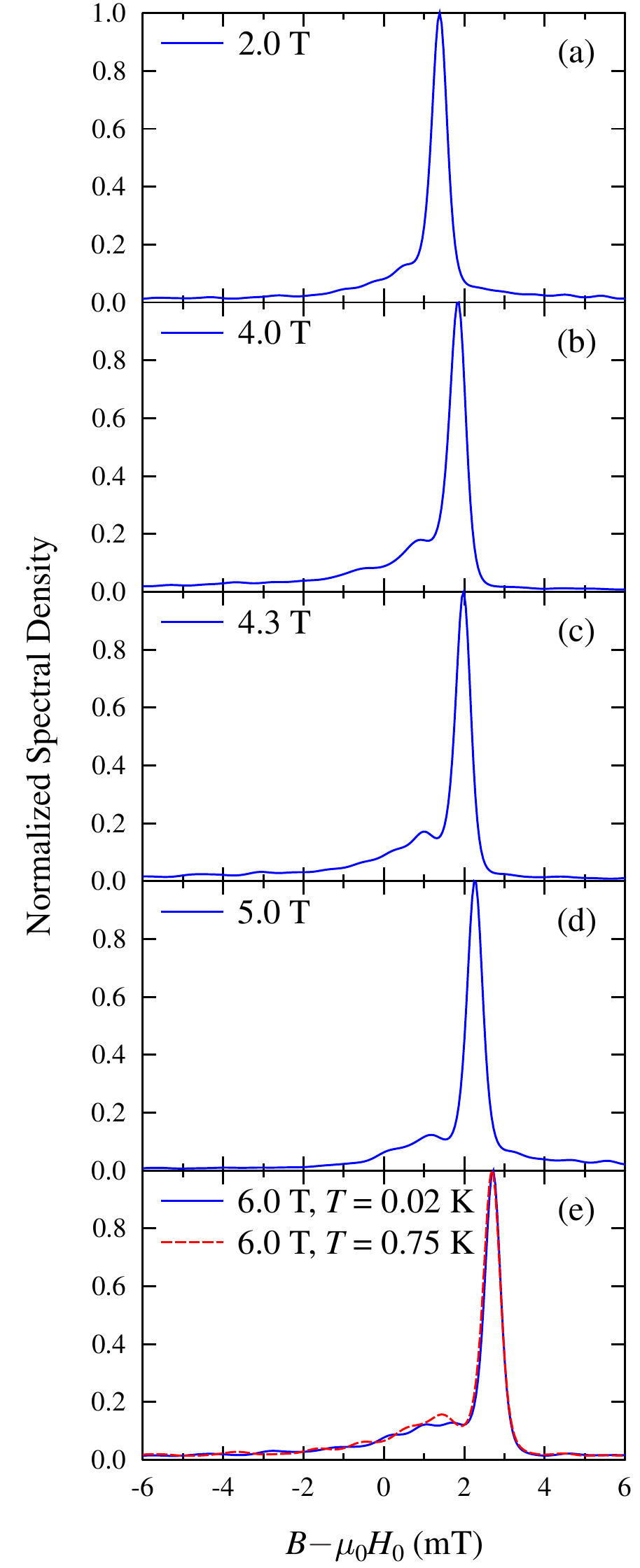}
	\caption{Fourier transform transverse-field $\mu^+$SR spectra measured at $T = 0.02~{\rm K}$. In (e) the corresponding spectrum at $T = 0.75~{\rm K}$ is shown as the dashed line.}
	\label{fig:maxent}
\end{figure}

To investigate the closure of the Haldane gap, transverse-field  $\mu^+$SR spectra were measured for a powder sample in applied fields $2.0\leq \mu_0 H_0 \leq 7.0~{\rm T}$ at fixed temperatures $T = 0.02~{\rm K}$ and $T=0.75~{\rm K}$. The distribution of  local magnetic field strength $p(B)$ across the sample volume may be obtained by taking the Fourier transform of the measured time-dependent muon spin polarization. Example Fourier spectra for $T = 0.02~{\rm K}$, obtained using the maximum entropy method \cite{rainford1994}, are shown in Figure~\ref{fig:maxent}. The spectra consist of a narrow main peak (common to all) and a low-field shoulder whose character changes with applied field $\mu_0 H_0$.  The main peak is due to muons experiencing fields close to $\mu_0 H_0$, in sites in the sample where there are not significant static internal local magnetic fields. This main peak shifts to slightly higher fields as the external field $\mu_0 H_0$ is increased, which could reflect a small (motionally narrowed) paramagnetic contribution to the total local field at these sites due to dilute end-chain spins.
The broader low-field feature is well coupled to the magnetic behavior, and reflects the contribution from muons experiencing an internal local field in addition to the applied field.  A significant local field at the muon site below $\mu_0 H_{\rm c}$ is unexpected, but we note that a high-field feature was observed in the $\mu^{+}$SR spectra of the molecular spin-ladder BPCB [\citenum{lancaster2018}] which was attributed to a muon-induced distortion at the muon site.  
While both the main peak and the low-field shoulder (and therefore the average spectral weight) shift to higher fields with increasing $\mu_0 H_0$, the shift is larger for the main peak, resulting in increasing separation of the peaks as $\mu_0 H_0$ increases.  At $\mu_0 H_0 \ge 6.0$~T [Figure~\ref{fig:maxent}(e)] the shoulder broadens significantly, which reflects the widening of the local magnetic field distribution as the field at the muon site increases.  The spectra measured at $T = 0.75~{\rm K}$ are similar to those measured at $T = 0.02~{\rm K}$ [see Figure~\ref{fig:maxent}(e)].

To parametrize the shapes of the field spectra we fit them to the sum of three Gaussian peaks, one to model the main peak, a second to capture the additional weight on the low-field side and a low amplitude, broad contribution to capture the tails of the main peak:
\begin{equation}
A(B)=\sum^3_{i=1} A_i \exp \left[- \frac{(B-B_i)^2}{2\sigma_i^2} \right].
\label{eqn1}
\end{equation}
The spectral weight (given by the product $A_i \sigma_i$) of the low-field component is approximately constant, with the peak amplitude decreasing as the feature broadens.  The low-field shoulder accounts for $\approx 25\%$ of the total spectral weight, with the two other components contributing approximately equally to the remaining spectral weight. As seen in Figure~\ref{fig:tf}, the width of the shoulder is approximately constant up to $\mu_0 H_0 \approx 4.0~{\rm T}$ beyond which the full width half maximum (FHWM) increases significantly with increasing $\mu_0 H_0$ for both $T=0.02~{\rm K}$ and $T=0.75~{\rm K}$.  Similar increases in the width of the field distribution have been seen in $\mu^{+}$SR experiments on the molecular magnet [Cu(pyz)(gly)](ClO$_4$) [\citenum{lancaster2014}] and in the candidate Bose-Einstein condensation material Pb$_2$V$_3$O$_9$ [\citenum{conner2010}], where rapid rises in the relaxation rate were observed at field-induced transitions.  This suggests that the system undergoes a crossover in behavior or phase transition for fields $\mu_0 H_0 \gtrsim 4.0~{\rm T}$. 

\begin{figure}[t]
	\includegraphics[width=0.75\columnwidth]{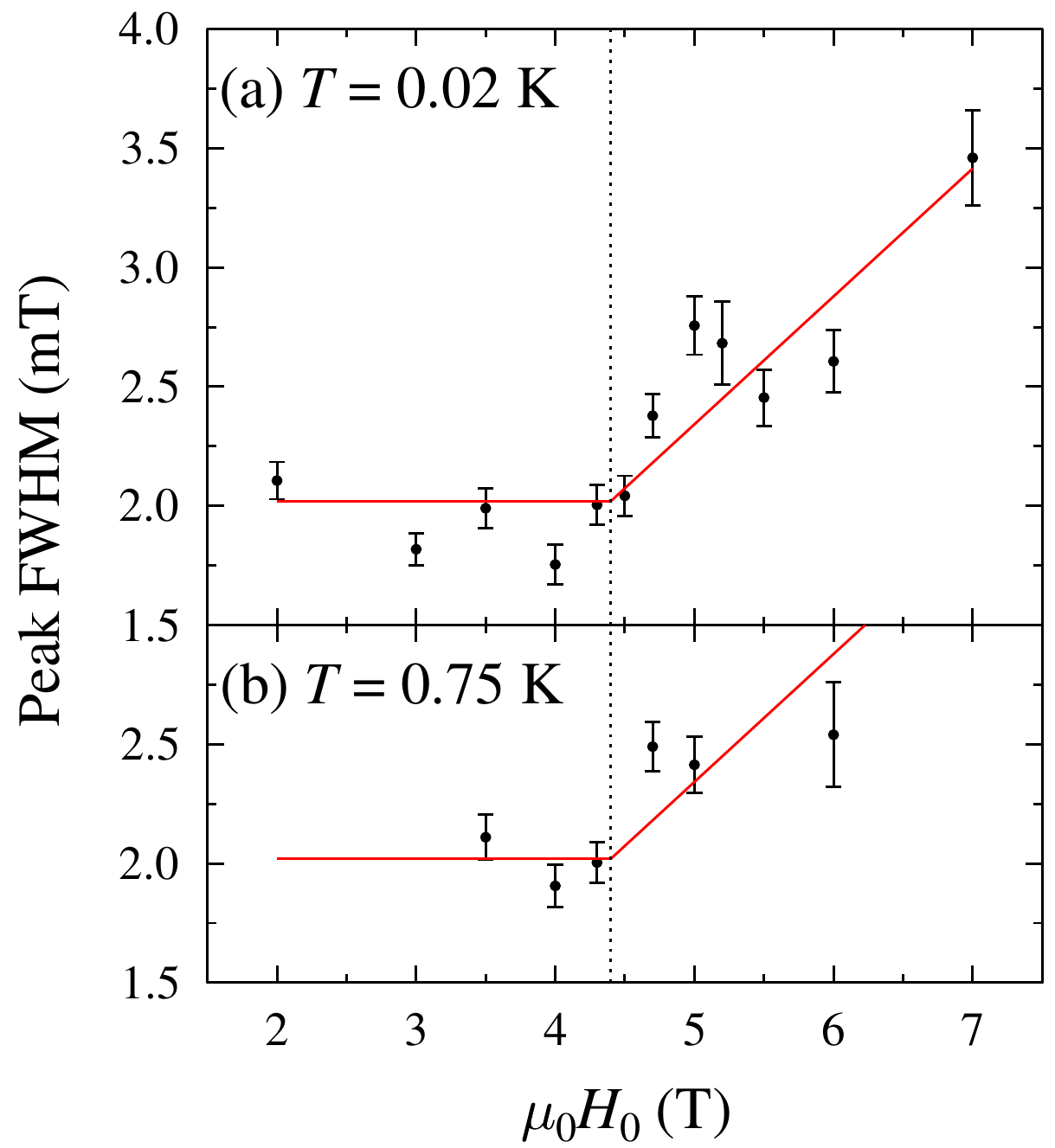}
	\caption{Full width half maximum (FWHM) for the low-field peak at (a) $T=0.02~{\rm K}$ and (b) $T=0.75~{\rm K}$. The solid lines are guides to the eye, and are the same for both temperatures.}
	\label{fig:tf}
\end{figure}

The solid line in Figure~\ref{fig:tf}(a) is a guide to the eye, capturing the field-dependence of the low-field peak FHWM for $T=0.02~{\rm K}$. We estimate $\mu_0 \bar{H}_{\rm c}\approx 4.4(2)~{\rm T}$ for the critical field, in agreement with the powder magnetization results. The same solid line is overlaid with the data for $T=0.75~{\rm K}$ in Figure~\ref{fig:tf}(b), showing that the behavior of the peak FHWM is similar at both temperatures.

\textbf{Electron Spin Resonance.}
In order to further elucidate the nature of the axial SIA induced by the Ni(II) tetragonal local environment, ESR measurements were performed on a powder sample of NiI$_2$(3,5-lut)$_4$. Transmission spectra collected in the first-derivative mode at high temperature are presented in Figure~\ref{fig:esr}(a), where $T = 30~{\rm K}$ is sufficiently high, relative to the energy scales of the intra-chain exchange coupling $J$ and resultant Haldane gap $\Delta_0$, that the system is within the thermally disordered phase. The spectra display a single exchange-narrowed transition and are therefore not indicative of the value of $D$. This has been seen elsewhere, for example in the related compound PbNi$_2$V$_2$O$_8$ [\citenum{smirnov2008}]. A linear fit to the transition positions in the field--frequency plane is shown as a dashed line in Figure~\ref{fig:esr}(a), which yields the powder-average $g$-factor $g=2.19(1)$, in close agreement with the value obtained using magnetic susceptibility.

\begin{figure}[ht]
  \includegraphics[width=\columnwidth]{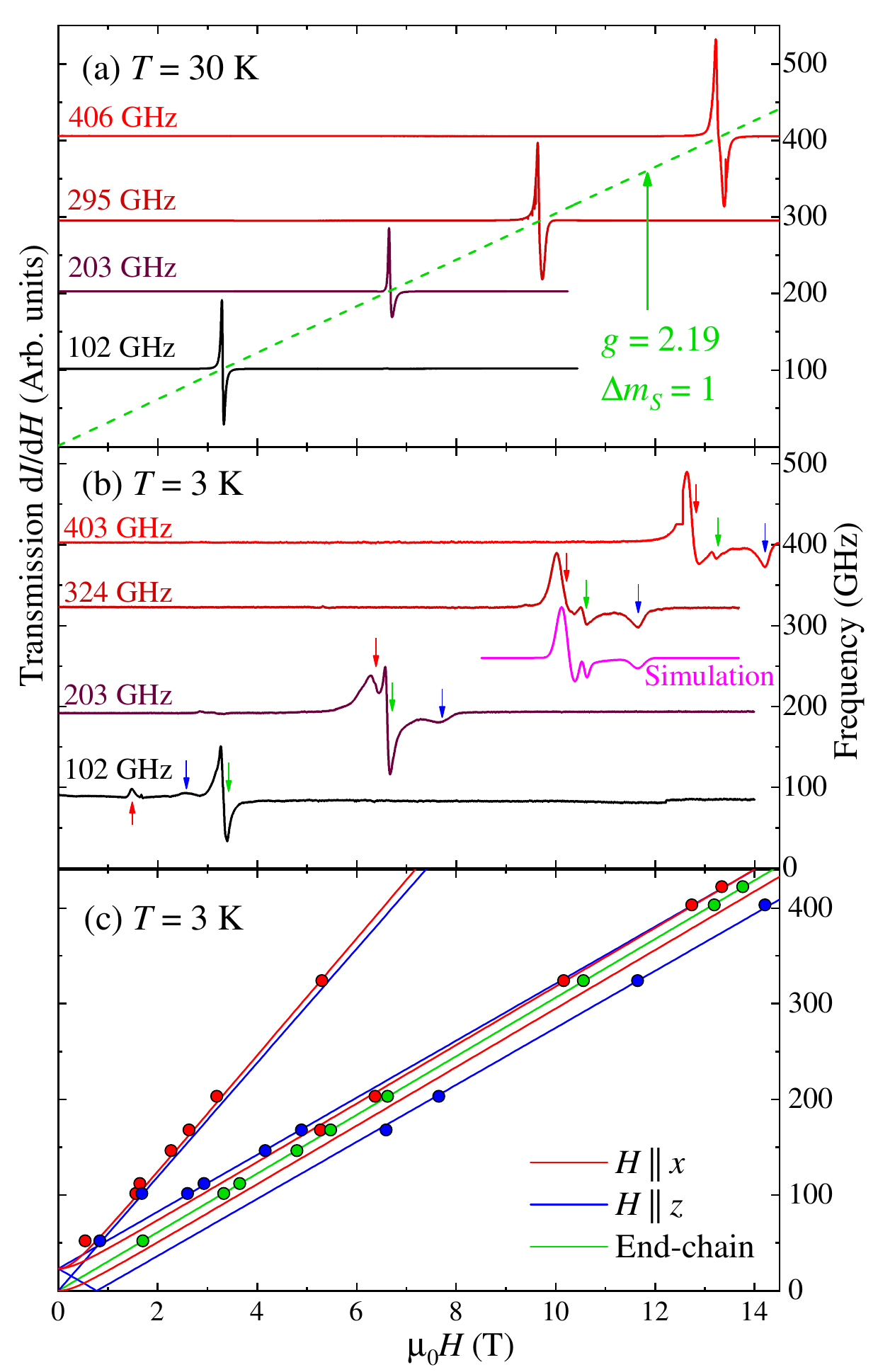}
  \caption{The frequency-dependence of ESR transmission spectra for a powder sample of NiI$_2$(3,5-lut)$_4$ at (a) $T=30~{\rm K}$ and (b) $T=3~{\rm K}$. (c) The positions of transitions observed at $T=3~{\rm K}$. The dashed line in (a) and solid lines in (c) are fits described in the text and the simulated spectrum in (b) is obtained with $\nu = 324~{\rm GHz}$.}
  \label{fig:esr}
\end{figure}

At low temperatures ($T = 3~{\rm K}$), the Haldane singlet phase is the ground state for applied magnetic fields below the anisotropic critical values $\mu_0 H_{\rm c}^{\parallel} = 5.3~{\rm T}$ and $\mu_0 H_{\rm c}^{\perp} = 4.3~{\rm T}$, whereupon a triplet state is driven lower in energy. ESR transitions between the singlet Haldane phase and triplet states are forbidden by momentum conservation, and are not observed in our data. 
The central feature which dominates at low frequencies [indicated by green arrows and circles in Figures~\ref{fig:esr}(b) and (c), respectively] is ascribed to fractional $S=1/2$ end-chain degrees of freedom, as observed in the magnetometry measurements, and is therefore considered separately to all other observed resonances.
This resonance is still observed for fields above the critical values, as seen in the Haldane compound NENB, for example \cite{cizmar2008}, albeit with a suppressed amplitude as the Haldane state depopulates for $H>H_{\rm c}$. The resonance positions are fit to an isotropic model for $S=1/2$ moments [green line in Figure~\ref{fig:esr}(c)], which yields the $g$-factor $g=2.191(4)$, consistent with the high-temperature result for the bulk $S=1$ moments.

The remaining resonances observed at $T=3~{\rm K}$ are indicated by red and blue arrows and circles in Figures~\ref{fig:esr}(b) and (c), and comprise satellite features at fields on either side of the central resonance, plus the `half-field' resonances corresponding to transitions with $\Delta m_S = \pm 2$ \cite{atherton}. At low frequencies and fields ($H \lesssim H_{\rm c}$) there are fewer transitions between excited triplet states, resulting in small amplitudes for the satellite peaks relative to the central end-chain resonance. 
However, for fields above the critical value a low-energy triplet state becomes the ground state, and these resonances rapidly gain intensity.

Based on their lineshapes \cite{atherton} (peak versus peak-derivative), the non-central transitions are ascribed to fields parallel to the $z$-direction (color-coded blue) and the $x$-direction (red). These positions may be fitted to the fermion model, which is equivalent to the description of isolated moments possessing axial SIA \cite{abragam}. This procedure captures the positions of the resonances,  as shown by the blue and red lines in Figure~\ref{fig:esr}(c), and yields the the anisotropic $g$-factors $g_{\parallel}=g_z=2.13(1)$ and $g_{\perp}=g_x=2.19(1)$, and an easy-plane \textit{effective} SIA parameter $D_{\rm eff}=+1.11(6)~{\rm K}$, and hence an easy-axis $D$ in the Hamiltonian. 
A simulated spectrum created by summing the two contributions outlined above is compared to the data measured for $\nu=324~{\rm GHz}$ in Figure~\ref{fig:esr}(b), where it may be seen that the lineshapes and positions of the main resonances (and the relative intensities of the two satellite peaks) are well described by the simulation.

\section{Discussion}

The INS and single-crystal magnetization studies provide direct measurements of the SIA-split Haldane gaps and anisotropic critical fields, respectively, for NiI$_2$(3,5-lut)$_4$. These quantities can be related via the model-dependent expressions given for the fermion and boson theories in Equations~\ref{eq:ferm} and \ref{eq:bos}, respectively. In order to test the validity of these theories for this system, the anisotropic $g$-factors that each model predicts, given the experimentally observed values of $\Delta^{\parallel , \perp}$ and $H_{\rm c}^{\parallel , \perp}$, are shown in Table~\ref{tab:g}. The two models yield an identical value of $g_{\parallel}=2.15(5)$ which is physically reasonable for Ni(II) ions, and in close agreement with the value determined using the low-temperature ESR measurements. The fermion  value of $g_{\perp}=2.21(6)$, and the resultant powder average $g_{\rm powder} = (g_{\parallel}+2g_{\perp})/3 =2.19(4)$ are also in excellent agreement with the high- and low-temperature ESR and powder susceptibility results. However, the boson  value of $g_{\perp}=1.85(6)$ is smaller than is plausible for this ion, and the powder average value $g_{\rm powder} = 1.95(4)$ is not in good agreement with the aforementioned techniques. 

\begin{table}[b]
\centering
\begin{tabular}{@{}l||c|c|c@{}}
Theory & $g_{\parallel} = g_z$ &  $g_{\perp}=g_{x}$ &  $g_{\rm powder}$ \\ \hline
Fermion  &  2.15(5) & 2.21(6) & 2.19(4)    \\ \hline
Boson     &  2.15(5) & 1.85(6) & 1.95(4)   \\ 
\end{tabular}
\caption{The anisotropic and powder $g$-factors predicted by the fermion (Equations~\ref{eq:ferm}) and boson (Equations~\ref{eq:bos}) models, using the values of $\Delta^{i}$ and $H_{\rm c}^{i}$ directly observed using INS and magnetization, respectively. }\label{tab:g}
\end{table}

\begin{figure}[t]
\centering
 \includegraphics[width=0.95\columnwidth]{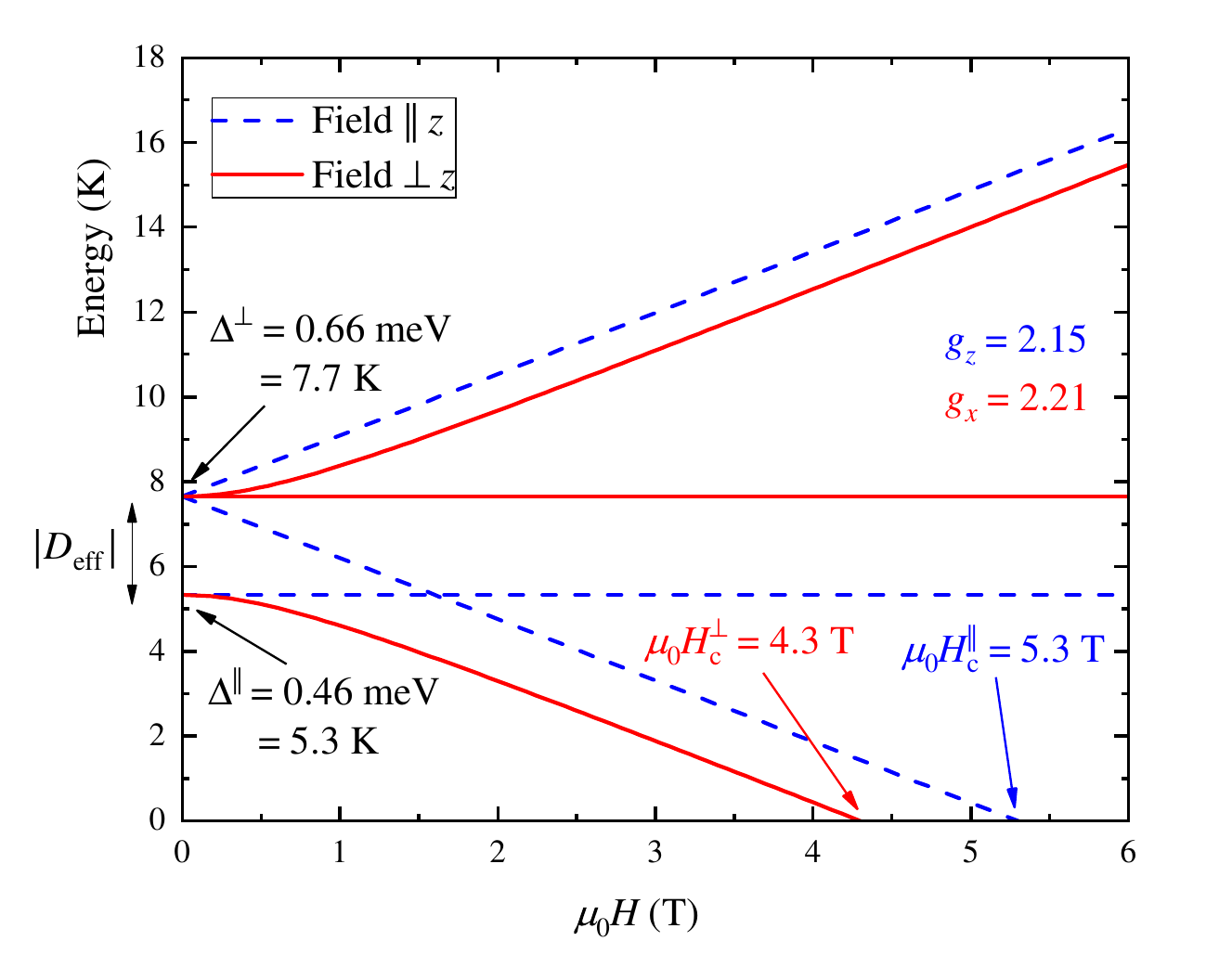}
 \caption{ The anisotropic field dependence of energy levels within the Haldane phase, according to the fermion model. The anisotropic critical fields and zero-field energy gap values  have been set to the values determined by magnetization (Figure~\ref{fig:SQUID2}) and INS (Figure~\ref{fig:ins}), respectively. The $g$-factors have been fixed to the values for the fermion model from Table~\ref{tab:g}, in order to satisfy the relations in Equation~\ref{eq:ferm}.  The effective SIA term $D_{\rm eff}=+2.4(2)~{\rm K}$ defined by Equation~\ref{eq:Deff} is labelled.}  \label{fig:energy}
\end{figure}

The fermion model is therefore in excellent agreement with the experimental results for the Haldane phase of NiI$_2$(3,5-lut)$_4$, demonstrating that its applicability is not limited to easy-plane Haldane compounds, as been recently suggested \cite{bera2015,bera2015a}. The resultant energy levels are shown in Figure~\ref{fig:energy}, where the values of $g_{\parallel , \perp}$ from Table~\ref{tab:g} reproduce the experimentally observed energy gaps $\Delta^{\parallel , \perp}$ and critical fields $H_{\rm c}^{\parallel , \perp}$. The  boson model has previously been successful in describing compounds with critical inter-chain interactions \cite{white2008,bera2015,bera2015a}, and  its failure to consistently describe the experimental findings for NiI$_2$(3,5-lut)$_4$ may constitute further evidence for small $|J_{\perp}|$ in this compound.

INS provides a \textit{direct}, zero-field measurement of the SIA-split triplet energy levels, and therefore unequivocally yields $D=-1.2(1)~{\rm K}$ (and $D_{\rm eff}=+2.4(2)~{\rm K}$), in excellent agreement with the estimate based on the anisotropic saturation behavior probed using pulsed-field magnetometry. Interestingly, the fermion model appears to describe powder ESR data for transitions between triplet states under the application of magnetic field, but with $D_{\rm eff} = +1.11(6)~{\rm K}$, which is close to half of the INS value. This discrepancy could be due to the fact that the energy levels of the Haldane chain are not well understood for fields above the critical field $H_{\rm c}$ [\citenum{kashiwagi2009}]. The fermion and boson models have been shown to fail in the description of energy levels above  $H_{\rm c}$, albeit in systems which undergo 3D magnetic ordering above the critical field. In the related Haldane compounds SrNi$_2$V$_2$O$_8$ and NDMAP,  INS [\citenum{bera2015a,zheludev2003,zheludev2004}] and ESR [\citenum{kashiwagi2009,hagiwara2003}] have demonstrated that the triplet energy levels change gradient as magnetic field is swept through $H_{\rm c}$, in contrast to the prediction of the fermion model.

\begin{table*}[t]
\centering
\begin{tabular}{@{}l||c|c|c|c|c|c|c|c|c@{}}
Compound & $J$~(K) & $D$~(K) & $D/J$ & $\Delta^{\parallel}$~(K) &  $\Delta^{\perp}$~(K) & $\Delta_0/J$ & $H_{\rm c}^{\parallel}$~(T) & $H_{\rm c}^{\perp}$~(T) & References  \\ \hline

NiI$_2$(3,5-lut)$_4$  &  17.5(2) & -1.2(1) & -0.07(1) & 5.3(1) & 7.7(1) & 0.40(1) & 5.3(1) &  4.3(1) & This study   \\ \hline

PbNi$_2$V$_2$O$_8$  &  104 & -5.2 & -0.05  & 21 &  28 & 0.25 & 19.0 & 14 & \cite{uchiyama1999,zheludev2000,smirnov2008}  \\ \hline

SrNi$_2$V$_2$O$_8$  &  101 & -3.7 & -0.04  & 18.2 &  29.9 & 0.26 & 20.8 &  12.0  & \cite{bera2013,bera2015} \\ \hline

Y$_2$BaNiO$_5$  &  280 & -9.4 & -0.03  & 87 & $100,110$ & 0.35 & \multicolumn{2}{c|}{$\sim 70$}  & \cite{sakaguchi1996,xu1996} \\ \hline

AgVP$_2$S$_6$     &  780 & 4.5 & 0.006 & \multicolumn{2}{c|}{$300$} & 0.38 & \multicolumn{2}{c|}{$\sim 200$}  & \cite{mutka1991} \\ 

\end{tabular}
\caption{A comparison of parameter values for selected Haldane compounds. }\label{tab:comp}
\end{table*}

The intra-chain exchange coupling $J = 17.5(2)~{\rm K}$ is deduced from the saturation behavior observed in the pulsed-field magnetization measurements, while the INS measurement yields $\Delta_0 = 7.0(1)~{\rm K}$. The resulting ratio $\Delta_0/J = 0.40(1)$ is very close to the value of $0.41$ expected for an ideal isolated chain, implying that the interchain exchange should not play a significant role in our material, at least at the temperatures explored here. In contrast, the SIA-split Haldane gaps in PbNi$_2$V$_2$O$_8$ and SrNi$_2$V$_2$O$_8$  are driven lower in energy by substantial inter-chain dispersion due to $J_{\perp}$ \cite{zheludev2000}.

Hence NiI$_2$(3,5-lut)$_4$ is remarkable as a highly isolated Haldane chain with a single crystallographic Ni(II) site, experimentally accessible energy scales ($J = 17.5(2)$~K, $\Delta^{\parallel,\perp} = 5.3, 7.7$~K, $H_{\rm c}^{\parallel,\perp} = 5.3, 4.3$~T) and an anisotropy ratio $D/J = -0.07(1)$. Systems with comparable degrees of anisotropy include Y$_2$BaNiO$_5$~\cite{xu1996}, PbNi$_2$V$_2$O$_8$ and SrNi$_2$V$_2$O$_8$~\cite{zheludev2000}, while AgVP$_2$S$_6$ is virtually isotropic~\cite{takigawa1995}. The parameters of all five compounds are compared in Table~\ref{tab:comp}. The large energy scales exhibited by both Y$_2$BaNiO$_5$ and AgVP$_2$S$_6$ preclude field-tuning through the quantum critical region associated with the closure of their Haldane gaps using DC magnetic fields. While $H_{\rm c}^\perp$ for PbNi$_2$V$_2$O$_8$ and SrNi$_2$V$_2$O$_8$ are within the range of superconducting magnets, both systems have near-critical values of $|J_{\perp}|/J$, lying on the very edge of the Haldane phase as evidenced by the departure of $\Delta_0/J$ from the ideal value of 0.41 and the observation of long-range magnetic order at liquid helium temperatures in applied fields. NiI$_2$(3,5-lut)$_4$ is therefore a uniquely ideal Haldane system that provides easy access to the entire Haldane phase for all field directions. In particular, the entire quantum critical region can be probed with techniques that may have hitherto not been viable for other compounds. This includes instruments at central beamtime facilities, as our preliminary $\mu^{+}$SR investigations demonstrate.

Furthermore, the molecular building blocks that make up NiI$_2$(3,5-lut)$_4$ permit a higher degree of controlled structural modification than is possible for the inorganic systems. Of particular interest is the controlled introduction of quenched bond disorder. Randomly distributed non-magnetic defects on the transition metal sites and substitutions which locally modify superexchange pathways have proven to be a pathway to exciting new physics \cite{zheludev2013}, yet still represent an unsolved problem. Controlled chemical substitution in related molecule-based materials has been demonstrated to give rise to entirely new phases, such as the elusive Bose-glass phase in the large-$D$ system DTN~\cite{yu2012} and the spin-ladder IPA-CuCl$_3$~\cite{hong2010a}. Bond disorder in the Haldane phase has been considered through numerical simulations, which predict localized impurity bound states \cite{sorensen1995,wang1996}. However, to date, the paucity of physically realized compounds where controlled introduction of bond disorder can be achieved has led to a lack of experimental verification of these theoretical predictions.

\section{Conclusions}

In summary, we have shown that NiI$_2$(3,5-lut)$_4$ is a realization of a $S=1$ Heisenberg AFM 1D chain, with intra-chain exchange coupling $J=17.5(2)~{\rm K}$ manifested by unique Ni--I$\cdots$I--Ni magnetic couplings. Muon spin relaxation measurements demonstrate the absence of zero-field magnetic LRO down to $T = 20~{\rm mK}$, indicating a high degree of one dimensionality (i.e.\ a small ratio $|J_{\perp}|/J$). The results of INS measurements prove this compound resides within the Haldane region of the theoretical phase diagram from Ref.~\cite{wierschem2014}, and quantify the Haldane energy gaps for the SIA-split triplet excitations $\Delta^{\parallel} = 0.46(1)~{\rm meV}$ and $\Delta^{\perp} = 0.66(1)~{\rm meV}$. These energy gaps correspond to an easy-axis SIA parameter $D = -1.2(1)~{\rm K}$, so that $D/J = -0.07(1)$, making this one of the most isotropic Haldane systems reported to date. Single-crystal magnetization measurements reveal the anisotropic critical fields $\mu_0 H_{\rm c}^{\parallel}=5.3(1)~{\rm T}$ and $\mu_0 H_{\rm c}^{\perp}=4.3(1)~{\rm T}$, which are readily accessible in commercially available superconducting magnets, and powder pulsed-field measurements demonstrate the system is saturated at $\mu_0 H_{\rm s}^{\parallel}=46.0(4)~{\rm T}$ and $\mu_0 H_{\rm s}^{\perp}=50.7(8)~{\rm T}$. These energy gaps and critical fields are explicable using the fermion model, in contrast to other reported easy-axis Haldane systems where $J_{\perp}$ is critically large and the boson model is more consistent with experiment \cite{smirnov2008,bera2015,bera2015a}.

NiI$_2$(3,5-lut)$_4$  is also ideally suited to the introduction of quenched disorder via substitution of the Ni(II) ion and the halide ions which mediate the superexchange interaction. This model system therefore provides a unique opportunity to explore this relatively uncharted territory and gain fresh insight into the physics of the Haldane phase using chemistry to control and tune the exchange interactions and SIA.

\section{Experimental Section}

\textbf{Synthesis.} All chemical reagents were purchased from commercial sources and used as received. A typical synthesis involves slow mixing of aqueous solutions of Ni(NO$_3$)$_2 \cdot 6$H$_2$O ($0.4002~{\rm g}$, $1.38~{\rm mmol}$) with excess NH$_4$I ($0.9971~{\rm g}$, $6.88~{\rm mmol}$). In 5-mL of acetonitrile, 2-mL of 3,5-lutidine was dissolved and slowly added by pipette to the aqueous Ni solution. The resulting solution is mostly blue with a slightly green hue. Slow solvent evaporation over the period of a few days resulted in yellow-green microcrystals. Larger single crystals can be grown using a vapor diffusion method wherein 3,5-lutidine is allowed to slowly diffuse into an aqueous solution containing Ni(NO$_3$)$_2 \cdot 6$H$_2$O and NH$_4$I. In either case, crystals of the product are obtained in better than 90\% yield. The crystals were found to be largely air-stable although they were generally stored in a refrigerator.

\textbf{Structural Determination.} Experiments were conducted on the ChemMatCARS 15-ID-B beamline of the Advanced Photon Source at Argonne National Laboratory. A microcrystal of NiI$_2$(3,5-lut)$_4$ measuring $10 \times 10 \times 2~\mu{\rm m}^3$ was selected from a bulk sample using a cryo-loop and mounted on a Huber 3-circle X-ray diffractometer equipped with an APEX II CCD area detector. The sample was cooled to $100(2)~{\rm K}$ using a LN$_2$ cryojet. Synchrotron radiation with a beam energy of $32.2~{\rm keV}$ ($\lambda = 0.41328$~\AA{}) was selected and the beam size at the sample position was $0.1 \times 0.1~{\rm mm}^2$. The distance between sample and detector was set at $60~{\rm mm}$. A total of $51,036$ reflections were collected of which $1,272$ were unique [$I > 2\sigma(I)$].
 
Data collection and integration were performed using the APEX II software suite. Data reduction employed SAINT.\cite{saint} Resulting intensities were corrected for absorption by Gaussian integration (SADABS).\cite{sadabs} The structural solution (XT)\cite{shelxt} and refinement (XL)\cite{shelxl} were carried out with SHELX software using the XPREP utility for the space-group determination. Considering systematic absences, the crystal structure was solved in the tetragonal space group $P4/nnc$ ($\#126$).\cite{itc} Lutidine H-atoms were placed in idealized positions and allowed to ride on the carbon atom to which they are attached. All non-hydrogen atoms were refined with anisotropic thermal displacement parameters.

\textbf{Magnetometry.} Quasi-static magnetometry measurements were performed using a Quantum Design SQUID magnetometer in fields up to $7~{\rm T}$, where an iQuantum $^{3}$He insert enabled temperatures $T \gtrsim 0.5~{\rm K}$ to be reached. The powder and crystal samples were loaded into a gelatin capsule, which was then attached to a nonmagnetic sample holder at the end of a rigid rod. The temperature-dependent molar magnetic susceptibility is obtained in the linear limit via $\chi_{\rm m} = M/(nH)$ where the powder contains $n$ moles of the compound.

Isothermal pulsed-field magnetisation measurements were performed at the National High Magnetic Field Laboratory in Los Alamos, USA. Fields of up to $65~{\rm T}$ with typical rise times $\approx 10~{\rm ms}$ were used, and a $^3$He cryostat provides temperature control. A powdered sample was mounted in a $1.3~{\rm mm}$ diameter PCTFE ampoule which was attached to a probe containing a 1500-turn, 1.5~mm bore, 1.5~mm long compensated-coil susceptometer, constructed from 50~gauge high-purity copper wire. When the sample is centered within the coil and the field is pulsed, the voltage induced in the coil is proportional to the rate of change of magnetization with time $({\rm d} M/{\rm d}t)$. The total magnetization is obtained by numerical integration of the signal with respect to time. A subtraction of the integrated signal recorded using an empty coil under the same conditions is used to calculate the magnetization of the sample. The magnetic field is measured via the signal induced within a coaxial 10-turn coil and calibrated using de Haas-van Alphen oscillations within the copper coils of the susceptometer. 

\textbf{Inelastic Neutron Scattering.} Zero-field inelastic neutron scattering measurements were performed on a powder sample of NiI$_2$(3,5-lut)$_4$ (mass $3.5~{\rm g}$) using the high-flux chopper setting on the direct geometry time-of-flight spectrometer LET, at the ISIS facility, UK. A primary incident energy $E_{\rm i} = 3.7~{\rm meV}$ was used, providing a number of incident energies with various flux and resolution combinations in the energy window of interest for this compound, of which we focus on the $E_{\rm i} = 2.2~{\rm meV}$ data. The sample was mounted in a $^4$He cryostat and datasets were collected at temperatures of $T = 1.7$, 3, 7 and $12~{\rm K}$, where the latter was treated as a non-magnetic background.

\textbf{Muon Spin Relaxation.} Zero-field $\mu^{+}$SR measurements on NiI$_2$(3,5-lut)$_4$ were made on the LTF and GPS instruments at the Swiss Muon Source (S$\mu$S), Paul Scherrer Institut, Switzerland.  Transverse-field $\mu^{+}$SR measurements were carried out on a polycrystalline sample using the HAL-9500 spectrometer at S$\mu$S. For the zero-field measurements, a powder sample was packed in Ag foil envelopes (foil thickness 12.5~$\mu$m), attached to a silver plate and mounted on the cold finger of a dilution refrigerator. For transverse-field measurements, a powder sample was packed in foil and glued to a silver holder that was mounted on the cold finger of a dilution refrigerator.  Data analysis was carried out using the WiMDA analysis program.\cite{pratt2000}

\textbf{Electron Spin Resonance.} High-field, high-frequency ESR spectra of powdered samples were recorded on a home-built spectrometer at the EMR facility, National High Magnetic Field Laboratory, Tallahassee, Florida, USA. Microwave frequencies in the range $52 \leq \nu \leq 422~{\rm GHz}$ and temperatures  $T = 3$ and $30~{\rm K}$ were used in the measurement. The instrument is a transmission-type device and uses no resonant cavity. A powdered sample was loaded into thin teflon vessel and lowered into a $^{4}$He cryostat. The microwaves were generated by a phase-locked Virginia Diodes source, with generating frequency $(13 \pm 1)~{\rm GHz}$, and equipped with a cascade of frequency multipliers to generate higher harmonic frequencies. The resultant signal was detected using a cold bolometer, and a superconducting magnet generated fields up to $15~{\rm T}$.

\section{acknowledgements}

We thank J. Liu for useful discussions. This project has received funding from the European Research Council (ERC) under the European Union's Horizon 2020 research and innovation programme (Grant Agreement No. 681260). WJAB thanks the EPSRC for additional support. A portion of this work was performed at the National High Magnetic Field Laboratory, which is supported by National Science Foundation (NSF) Cooperative Agreement No. DMR-1157490, the State of Florida and the U.S. Department of Energy (DOE) through the Basic Energy Science Field Work Proposal ``Science in 100 T''. Work at EWU was supported by NSF grant No. DMR-1703003. NSF's ChemMatCARS Sector 15 is principally supported by the Divisions of Chemistry (CHE) and Materials Research (DMR), NSF, under grant number NSF/CHE-1834750.  Use of the Advanced Photon Source, an Office of Science User Facility operated for the U.S. DOE Office of Science by Argonne National Laboratory, was supported by the U.S. DOE under Contract No. DE-AC02-06CH11357. JAS acknowledges support from the Independent Research/Development (IRD) program while serving at the National Science Foundation. FX acknowledges funding from the European Union's Horizon 2020 research and innovation program under the Marie Skodowska-Curie grant agreement No 701647. A portion of this work was performed at the Swiss Muon Source, Paul Scherrer Institute, Switzerland and we acknowledge support from EPSRC under grants EP/N023803/1, EP/N024028/1 and EP/N032128/1. BMH thanks STFC for support via a studentship. Data presented in this paper resulting from the UK effort will be made available at (URL here).

\bibliography{haldane}

\end{document}


\title{Supporting Information accompanying\\
A Near-Ideal Molecule-Based Haldane Spin-Chain}

\pacs{}
\maketitle

\begin{figure}[b]
	\centering
	\includegraphics[width=\columnwidth]{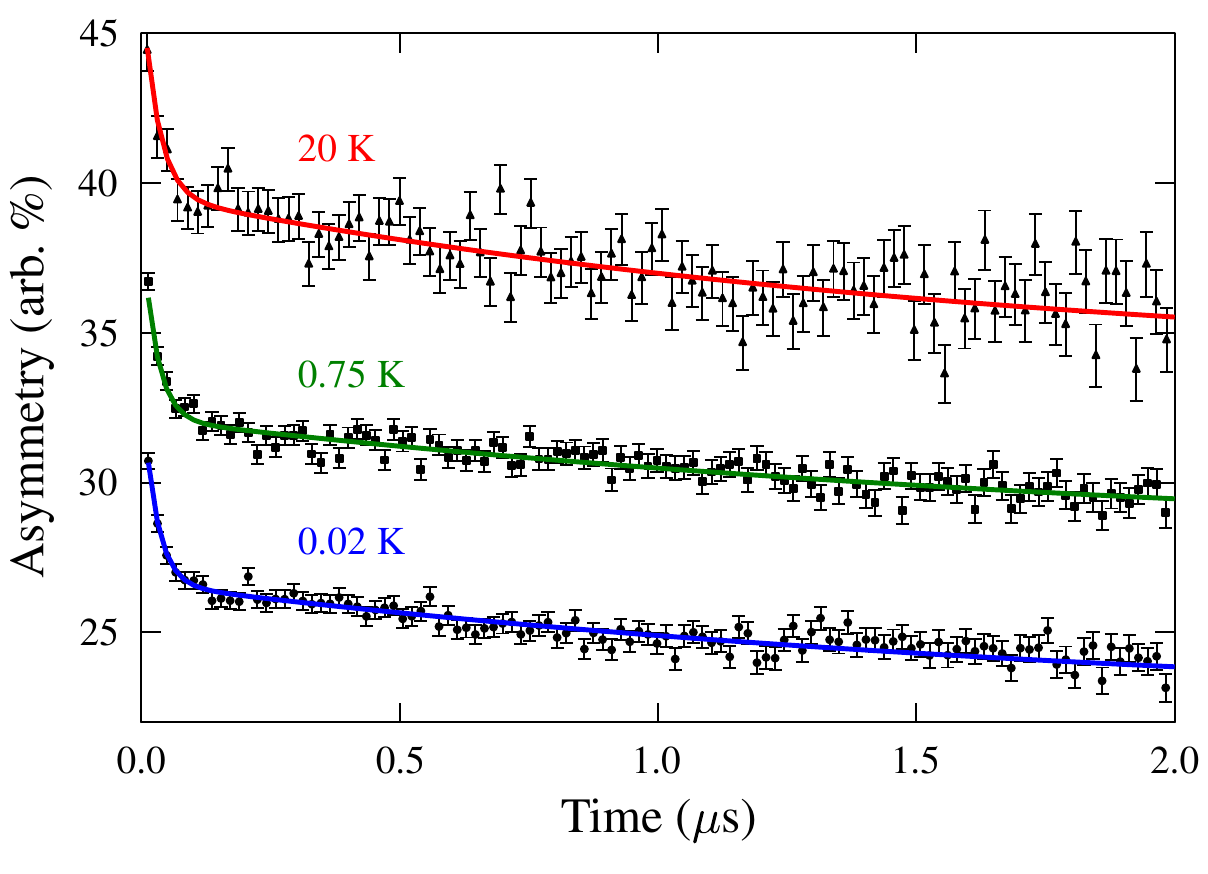}
	\caption{Asymmetry from ZF $\mu^{+}$SR measurements on NiI$_2$(3,5-lut)$_4$ at three different temperatures. (The spectra are offset for clarity.) The solid lines are fits to Equation~\ref{eqn:fit}.}
	\label{fig:fits}
\end{figure}

In a muon spin relaxation ($\mu^+$SR) experiment, a beam of spin-polarized positive muons is incident upon a material \cite{blundell1999,yaouanc}. Muons stop within the sample at particular sites, where their spins precess in the local magnetic field, before decaying after, on average, $2.2~\mu{\rm s}$. The observed property of the experiment is the average spin polarization of muon ensemble. In a zero-field (ZF) $\mu^+$SR experiment the muon spins precess entirely in the internal magnetic field of the sample. If a material shows quasistatic LRO, we expect spontaneous coherent oscillations in $P_z(t)$, the longitudinal spin polarization along the initial muon spin direction. In the TF $\mu^+$SR experimental geometry \cite{blundell1999} the externally applied field $\mu_0{\bm H}_0$ is directed perpendicular to the initial muon spin direction. Muons precess about the total magnetic field ${\bm B}$ at the muon site, which is a vector sum of $\mu_0{\bm H}_0$ and fields due to the Ni(II) moments. The observed property of the experiment is the time evolution of the muon spin polarization $P_x(t)$. The Larmor relation between muon precession frequency and local magnetic field strength $\omega = \gamma_{\mu}B$ (where $\gamma_{\mu}/(2 \pi) = 135.5~{\rm MHz~ T}^{-1}$ is the muon gyromagnetic ratio) allows the determination of the distribution $p(B)$ of local magnetic fields across the sample volume by means of a Fourier transform of $P_x(t)$.  

\textbf{Zero-field $\mu^+$SR measurements.} Example ZF spectra spanning the measured temperature range for  a powder sample of NiI$_2$(3,5-lut)$_4$  are shown in Figure~\ref{fig:fits}. The asymmetry data were fitted in the time domain with two exponentially decaying components and a constant background $A_\text{bkg}$,
\begin{equation}\label{eqn:fit}
	A\left(t\right) = A_1e^{-\lambda_1t} + A_2e^{-\lambda_2t} + A_\text{bkg} ,
\end{equation}
where $\lambda_i$ are relaxation rates. The two separate exponentials capture both fast-relaxing and slow-relaxing components of the data. The fast relaxing component typically arises from muons in bound states \cite{lancaster2006} whereas the slow relaxing component likely arises from fluctuations of magnetic fields detected by diamagnetic muon states, and is therefore sensitive to the magnetism of the system. The relaxation due to muon bound states is not expected to change with temperature, and as the value of $\lambda_2 \simeq 38$~MHz remained roughly constant over the whole temperature range, it was fixed at this value for the fits. Similarly, the ratio of muons in bound states to those in diamagnetic states should remain constant with temperature (as the muon site is determined by the structure of the material). Approximately constant values of $A_1$ and $A_2$ were found over the whole temperature range, as expected, and hence $A_1$ and $A_2$ were also fixed.

\begin{figure}[t]
	\centering
	\includegraphics[width=\columnwidth]{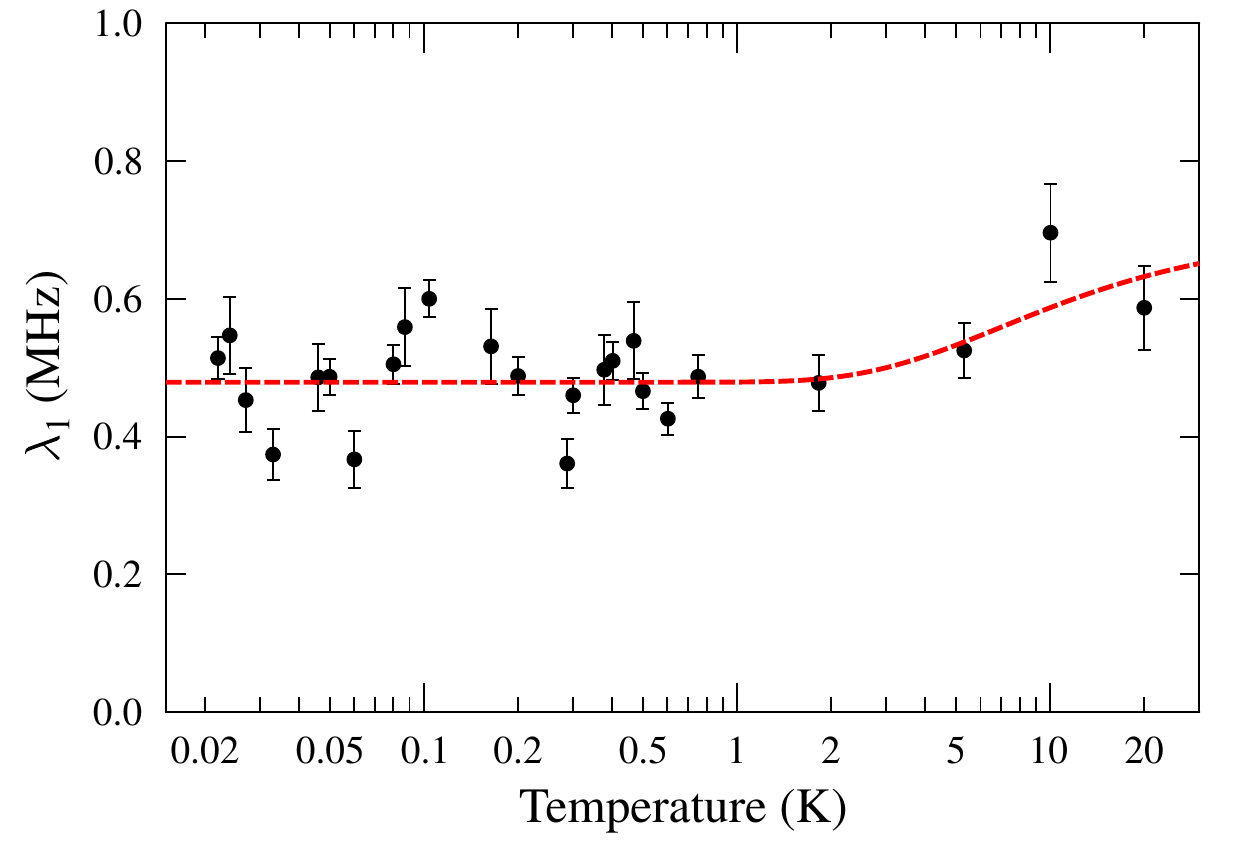}
	\caption{Values of $\lambda_1$ obtained by fitting ZF $\mu^+$SR data to Equation~\ref{eqn:fit}. Red dashed line shows $\lambda_{1}  \propto \exp\left(-\Delta/k_{\rm B} T\right)$ with $\Delta = 7$~K.}
	\label{fig:lambda}
\end{figure}

The value of $\lambda_1$ as a function of temperature is shown in Figure~\ref{fig:lambda}. At a phase transition, one would expect a dramatic increase in the value of $\lambda_1$, indicating that there is a large variation in the values of internal magnetic fields, or a slowing of dynamics. As the value of $\lambda_1$ stays approximately constant throughout the whole temperature range this suggests there is no phase transition,  indicating that there is no transition to long-range order even down to the lowest measured temperatures.

Previous work \cite{capponi2019} has shown that the relaxation rate $\lambda_{1}$ in a Haldane chain should vary as $\propto \exp\left(-\gamma\Delta/k_{\rm B} T\right)$, where $\Delta$ is the gap magnitude and the value of the constant $\gamma$ depends on the temperature regime and details of the model. Figure~\ref{fig:lambda} shows this scaling law for $\Delta=7$~K and $\gamma = 1$ (expected for low temperatures), suggesting that the slight increase in $\lambda_{1}$ above the gap temperature is consistent with the predicted behaviour. Although there is insufficient data density to test which value of $\gamma$ best describes this system, the location of the upturn is supportive of the value $\Delta \approx 7 K$ found through other techniques.

There are several other possible causes of the occurrence of the relaxation parametrized by $\lambda_{1}$, all consistent with the realisation of a Haldane phase in this material. One possibility is that chain ends give rise to magnetic moments, which introduce some magnetic field dynamics that cause the relaxation. Another possibility is that defects in the sample affect the local field configuration, again leading to a small, fluctuating magnetic field. It is also possible that the muon itself distorts the local environment at the implantation site inducing a small local moment. This has been observed previously in the magnetically disordered state of molecular spin-ladder systems \cite{lancaster2018}, where it was shown that it did not prevent the muon being a faithful probe of the intrinsic physics of the system.

\bibliography{haldane_SI}